\documentclass[letterpaper]{article} 
\usepackage{aaai2026}  
\usepackage{times}  
\usepackage{helvet}  
\usepackage{courier}  
\usepackage[hyphens]{url}  
\usepackage{graphicx} 
\usepackage{natbib}  
\usepackage{caption} 
\usepackage{algorithm}
\usepackage{algorithmic}
\usepackage{multirow}
\usepackage{newfloat}
\usepackage{listings}
\usepackage{booktabs}
\usepackage{makecell}
\usepackage{amsmath, amsfonts}
\usepackage{dsfont}

\DeclareCaptionStyle{ruled}{labelfont=normalfont,labelsep=colon,strut=off} 
\floatstyle{ruled}
\newfloat{listing}{tb}{lst}{}
\floatname{listing}{Listing}
\title{Chinese Court Simulation with LLM-Based Agent System}

\author{
    Kaiyuan Zhang\textsuperscript{\rm 1}\thanks{ky-zhang24@mails.tsinghua.edu.cn}
    Jiaqi Li\textsuperscript{\rm 1}
    Yueyue Wu\textsuperscript{\rm 1}\thanks{Corresponding Author: wuyueyue1600@gmail.com}
    Haitao Li\textsuperscript{\rm 1}
    Cheng Luo\textsuperscript{\rm 2}
    Shaokun Zou\textsuperscript{\rm 1}
    Yujia Zhou\textsuperscript{\rm 1}
    Weihang Su\textsuperscript{\rm 1}
    Qingyao Ai\textsuperscript{\rm 1}\thanks{Corresponding Author: aiqy@tsinghua.edu.cn}
    Yiqun Liu\textsuperscript{\rm 1}
}
\affiliations{
    \textsuperscript{\rm 1}Department of Computer Science and Technology, Tsinghua University\\
    \textsuperscript{\rm 2}MegaTech.AI Inc.

}

\usepackage{bibentry}

\begin{document}

\maketitle

\begin{abstract}
Mock trial has long served as an important platform for legal professional training and education. It not only helps students  learn about realistic trial procedures, but also provides practical value for case analysis and judgment prediction. Traditional mock trials are difficult to access by the public because they rely on professional tutors and human participants. Fortunately, the rise of large language models (LLMs) provides new opportunities for creating more accessible and scalable court simulations. While promising, existing research mainly focuses on agent construction while ignoring the systematic design and evaluation of court simulations, which are actually more important for the credibility and usage of court simulation in practice. To this end, we present the first court simulation framework -- SimCourt -- based on the real-world procedure structure of Chinese courts. 
Our framework replicates all 5 core stages of a Chinese trial and incorporates 5 courtroom roles, faithfully following the procedural definitions in China. 
To simulate trial participants with different roles, we propose and craft legal agents equipped with memory, planning, and reflection abilities. Experiment on legal judgment prediction show that our framework can generate simulated trials that better guide the system to predict the imprisonment, probation, and fine of each case.
Further annotations by human experts show that agents’ responses under our simulation framework even outperformed judges and lawyers from the real trials in many scenarios. 
These further demonstrate the potential of LLM-based court simulation. 
\end{abstract}

\begin{links}
    \link{Code} https://github.com/Miracle-2001/SimCourt
    \link{Datasets} https://github.com/Miracle-2001/SimCourt
\end{links}

\begin{figure}[t]
\centering 
\includegraphics[width=1\linewidth,keepaspectratio]{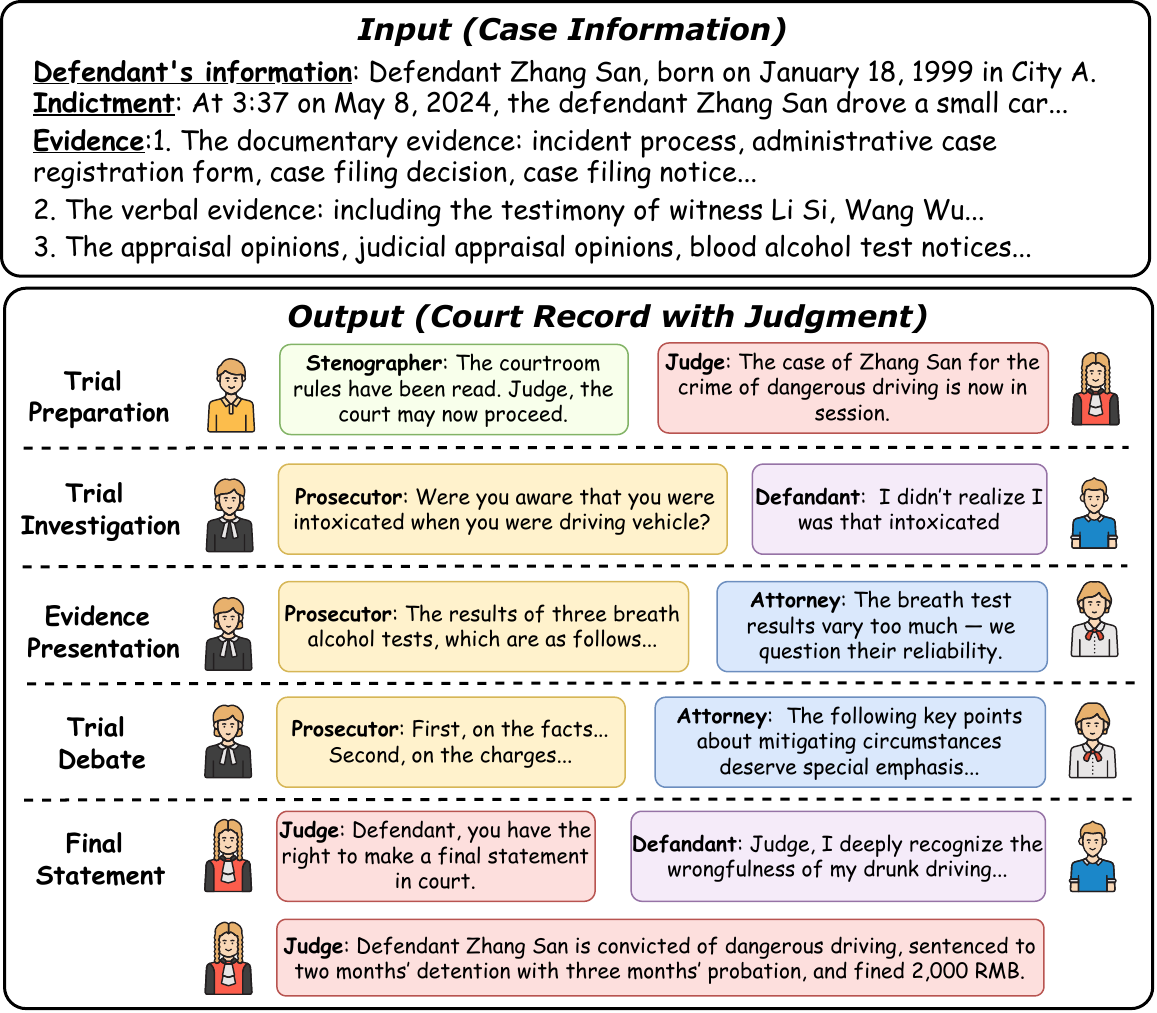}
\caption{The input of SimCourt consists of 3 components: defendant's information, indictment and evidence. Its output is a complete 5-stage trial proceeding record, culminating in a final judgment.}
\label{input-output}
\end{figure}

\section{Introduction}


Legal education and training play a critical role in cultivating professional skills and fostering a sense of justice among future legal practitioners. Mock trial, a widely adopted educational method, enables students and professionals to simulate trial proceedings, sharpen courtroom reasoning, and gain hands-on legal experience \cite{issa2023impact, tang2021effective}. However, traditional mock trials heavily rely on expert participation, proper venues, and offline human involvement. Additionally, they are time-consuming, expensive, and lack reproducibility \cite{zhang2021construction}. These limitations constrain the scalability of the traditional mock trial. 






Recent advances on large language models (LLMs) and LLM-based agents bring new opportunities. LLMs are capable of instruct following, reasoning, text generation, and human socialization understanding \cite{zhao2023more}. With carefully designed modules and external tools, LLM-based agents can perform as intelligent tools for specific domain \cite{xi2023rise,zhang2024survey}. These capabilities are highly compatible with the demands of rule adherence, courtroom debate, long-context response and legal reasoning inherent in judicial proceedings. By constructing LLM agents to simulate different roles in a courtroom, we can conduct court simulation automatically, offering valuable training opportunities and references for legal professionals.


 
Despite this potential, existing studies on court simulation remain limited in scope. They typically focus on simplified debate segments that fail to cover the full procedural of a trial, and there isn't a well-established and rigorous evaluation framework of both the judgment accuracy and interaction quality in the simulated trials. These limitations severely restrict their applicability in legal education and decision. In addition to procedural design and rigorous evaluation, further efforts are also needed in the agent system design. While powerful, general-domain LLMs still remain limited in maintaining courtroom roles, generating coherent arguments and making logical legal reasoning under courtroom scenarios. Therefore, it is essential to propose a structured, qualified agent system in which all critical roles are well simulated.

To address the aforementioned problems, we proposed a novel mock trial framework-- SimCourt -- grounded in the official procedural structure of criminal trials in Chinese courts. As illustrated in Figure.\ref{input-output}, the input of SimCourt consists of basic case materials collected before a trial, including the defendant's information, the indictment and corresponding evidences. The output of the simulation is a full 5-stage trial record with corresponding judgment documents, which contains the final sentence and explanations created by the judge. In order to achieve professional, coherent, and logical court simulation, we carefully designed different courtroom agents with three modules: the Profile, Memory, and Strategy module. The Profile module helps the agent understand its role and goal, the Memory module tracks the progression of the trial, and the Strategy module guide the agent to develop plans and generate responses accordingly. To compensate agents built from general LLMs with domain-specific knowledge, each agent is equipped with legal retrievers that can retrieve legal articles and relevant cases. This design helps us to simulate different courtroom roles under the same agent structure.

\begin{figure*}[]
\centering 
\includegraphics[height=6.3cm,keepaspectratio,width=1\linewidth]{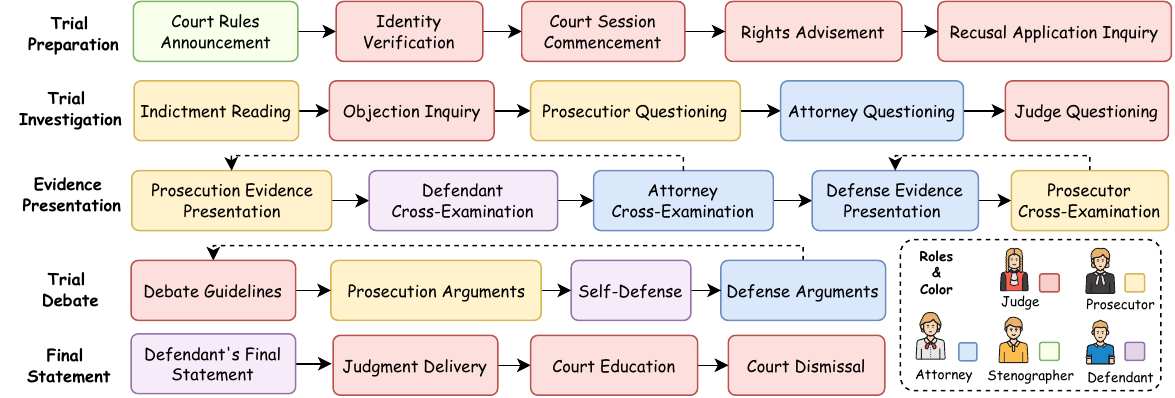}
\caption{The flowchart of the trial procedure in SimCourt, which is grounded on real trial procedures. Colors of the boxes represent the roles that lead the respective phase. The correspondence between roles and colors can be found at the bottom right corner of the chart. In the diagram, the dashed arrows represent optional backtracking.} 
\label{flowchart}
\end{figure*}

To comprehensively evaluate the quality of our LLM-based mock trial, we develop an evaluation framework as well as a benchmark based on both the objective goals of the court and a process evaluation guideline widely used by Chinese legal professionals\footnote{\url{http://gongbao.court.gov.cn/Details/ee6a5b1d20140c38c800c91c728d63.html}}. The framework consists of two parts, namely the evaluation of judgment prediction and the evaluation of court simulation process.
In judgment prediction, we extract the final sentences from the output of SimCourt and compare it with the real sentences recorded for the same trial.
In process evaluation, we summarize 30 aspects commonly used by legal experts to judge the performance of a judge or a lawyer, and conduct pairwise annotations to compare the quality of responses generated by our system and real human professionals in the trials.
For better reproducibility, we construct a benchmark for mock trial based on the above evaluation framework by randomly sampled 200 criminal case documents and 20 video recordings from the most common 40 criminals
from LeCard2.0 dataset \cite{li2024lecardv2} and China Court Trial Online\footnote{\url{https://tingshen.court.gov.cn/}}, respectively.
Experiment results show that SimCourt can generate high-quality trial recordings and agents in our system outperform real human in court from many perspectives.






The main contributions of our work can be summarized as followings:

(1) We proposed SimCourt, a Chinese criminal court simulation framework based on real trial procedures.

(2) We designed a courtroom agent framework with memory, planning and reflection abilities. Equipped with multiple modules and external legal tools, agents can simulate professional lawyers or judges and generate high-quality trial records in SimCourt. 



(3) We develop a comprehensive evaluation framework for LLM-based mock trial based on the quality of judgment prediction and a process evaluation guidelines widely used by Chinese judges. Experiment results show that our system and agents can generate responses better than those recorded in real trials from many perspectives, demonstrating the potential of SimCourt for legal practice and education.

\section{Related Work}

\textbf{Framework of LLM-based agents} Agents powered by LLMs can present human-like actions and has potential opportunities in areas such as role-playing, social simulation and code generation \cite{guo2024large}. Generally, a LLM-based agent should have a perception and action module to interact with the environment \cite{xi2023rise}. To better achieve typical goals, agents should also be equipped with memory module, external knowledge base and planning ability \cite{zhang2024survey}. Approaches including reason-action combination and tool usage can also enhance the performances of agents on various tasks \cite{yao2023react,qin2024tool}. Different from internal mechanism design in single agent system, a multi-agent system emphasizes diverse agent profiles, inter-agent communications, and collective decision-making abilities \cite{guo2024large}.





\textbf{World Simulation with LLM-based agents}. Leveraging the preliminary understanding of the real world of LLMs, researchers have recently pioneered a new direction: simulating the real world using agents powered by LLMs \cite{mou2024individual}. For instance, Wang et.al.\cite{wang2023recagent} examine the recommendation system via multi-agent communication on a virtual social media platform. Li et.al. \cite{li2024agent} construct an entire hospital treatment progress from registration to doctor consultant with LLM-based agent systems. Further researches on software development, gaming, employment consultant and social science area also validate novel application values of LLM-based agent system on real world simulation \cite{qian2023chatdev,xu2023exploring,tang2024gensim,xie2024can,jin2024agentreview}. 


For court simulation area, Chen et.al. provides a simplified trial debate simulation framework named AgentCourt~\cite{chen2024agentcourt}, in which lawyer agents can evolve with adversarial debating. He et.al.~\cite{he2024agentscourt} proposed AgentsCourt, a judicial decision-making agent with trial debates simulation and legal knowledge augmentation, which shows the importance of constructing trial process in legal judgment prediction (LJP) task. However, comparing to SimCourt, these works ignored the systematic process design and evaluation of simulations, resulting in restricted application value in legal education and decision support.








\section{Methodology}


This section presents the overall structure of SimCourt and the design of agents for simulating different courtroom roles.

\paragraph*{Trial Procedure and Participants}

The overall procedure and participants of our simulated court in SimCourt is presented in Figure.\ref{flowchart}.
SimCourt is grounded in the official procedural structure of criminal trials in Chinese courts with 5 stages (i.e., Trial Preparation, Trial Investigation, Evidence Presentation, Trial Debate and Final Statement) and 5 critical roles (i.e., Judge, Prosecutor, Attorney, Defendant and Stenographer) involved. The detailed explanations of each stage are provided below: 



\textbf{Trial Preparation Stage} represents the start of a trial. Here, Stenographer first announces the rules of the court. Then Judge in turn verifies the identity of parties, advise the Defendant's rights and make inquires about recusal. 

\textbf{Trial Investigation Stage} focuses on verifying the basic facts about the case. 
In this stage, Prosecutor and Attorney in turn examine the Defendant about the case proceedings, outcomes, post-case handling and other related aspects. Judge may also choose to questioning the defendant if necessary. 

\textbf{Presentation of Evidence Stage} is the time when Prosecutor and Attorney present their evidence in sequence. 
The evidence is subject to examination by the opposing party. 

\textbf{Trial Debate Stage} is one of the most important part of a trial. Prosecutor and Attorney engage in a comprehensive debate based on their respective positions, covering the nature of the case, the application of laws, and sentencing recommendations. Judge listens to both parties' viewpoints and provide guidance at appropriate times. 

\textbf{Final Statement Stage} is where Judge grants Defendant the right to make a final allocution for fairness and justice. After that, Judge issues the final judgment documents based on the case details and trial arguments.

To navigate and finish each stage of the courtroom process smoothly and effectively, each participant of the court must perform their jobs accurately and professionally. This means that participants need to not only produce contextually coherent, logical, and flexible utterances, but also behave with certain level of legal proficiency. Specifically, \begin{itemize}
 \item \textbf{Defendant} needs to protect his or her own interests within the scope of the case information. 
 \item \textbf{Prosecutor} should hold a proper questioning strategy, make logical accusations and effective defense rebuttal. 
 \item \textbf{Attorney} should be aware of the Defendant's rights, and propose logical evidence rebuttal and defense arguments. 
 \item \textbf{Judge} must ensure the fairness in the courtroom, neutrally control the procedural, and make interventions or guidance when necessary. Also, he or she must fully investigate the truth and issue appropriate judgment based on the case details and trial arguments, typically including imprisonment, possible probation, and fines. 
 \item \textbf{Stenographer} should record the trial and participants' interactions faithfully and correctly.
\end{itemize}

\paragraph*{Agent Structure and Implementation}

\begin{figure}
    \centering
    \includegraphics[width=\linewidth]{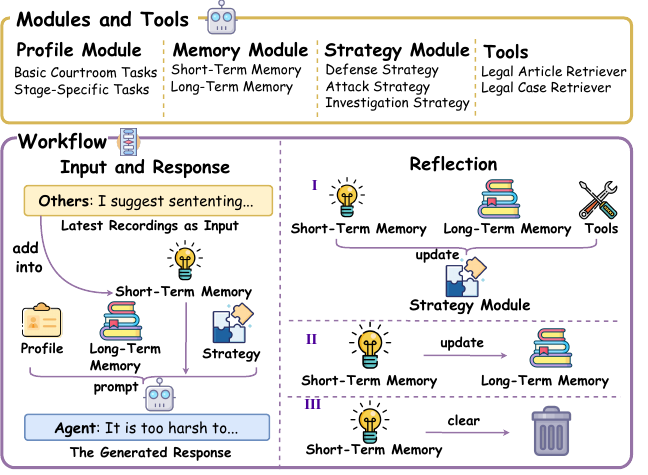}
    \caption{Agent structure and workflow of SimCourt, which consists of Profile module, Memory module, Strategy module and two external tools. The input-response and reflection workflow is also illustrated.}
    \label{fig:agent}
\end{figure}

For carrying out the jobs of the participants, the agent we proposed mainly consists of 3 modules and 2 external tools, as shown in Figure.\ref{fig:agent}. Detailed descriptions on modules, tools and other mechanisms are introduced in this section.


\textbf{Modules and Tools}.
Profile, Memory, and Strategy modules are designed separately to guide the agent to conduct effective role-playing in the courtroom.
The Profile module consists of the descriptions of basic tasks for courtroom roles and stage-specific tasks for each specific stage. It guides the agent to maintain the characteristics of their specific roles and follow the trial procedure accurately. 
The Memory module is constructed with a short-term memory and a long-term memory.
The short-term memory memorizes the latest statements and interactions in the current stage, and the long-term memory uses LLMs to summarize critical statements of previous stages and then store them for later usage.
Both short-term and long-term memory are fed into LLM agents to generate the output. The Strategy module is designed to maintain a consistent debate and negotiating strategy for Defendant, Prosecutor, Attorney, and Judge.
It assists the agents to generate structured, logical, and appropriate statements. 
Specifically, there are 3 types of strategies implemented in SimCourt, which are Defense strategy, Attack strategy, and Investigation strategy. The Defense strategy focuses on maintaining the relevance and probative value of one's opinions; the Attack strategy focuses on undermining the opposing party's opinions; and the Investigation strategy focuses on clarifying critical issues related to conviction and sentencing.
Each agent create their strategies based on the case information and their profile at the beginning of the trial, and use these strategies to guide their behaviors.

To compensate general LLMs with legal domain knowledge, we further include two useful tools, i.e., a legal article retriever and a legal case retriever, in each courtroom agent. The legal article retriever can retrieve articles from a law base consists of 55,347 Chinese laws, including all articles in the Criminal Law of the People's Republic of China. Here we adopt the simplest yet effective approach by directly retrieving the content of legal provisions based on their names and numbers. 
The legal case retriever is implemented with a commercial search service named LegalOne\footnote{\url{https://legalone.com.cn/}}, which is a LLM-based search engine that can retrieve and summarize judicial cases according to natural language queries. LegalOne is built from a large-scale corpus with more than 2 million legal documents and 200,000 representative cases selected by The Supreme People’s Court of China.

\renewcommand{\arraystretch}{1.1}

\begin{table*}[]
\centering
\small
\begin{tabular}{c|lc|lc|lc}
\hline
\multirow{2}{*}{\textbf{Method}} & \multicolumn{2}{c|}{\textbf{Imprisonment}}                        & \multicolumn{2}{c|}{\textbf{Probation}}                           & \multicolumn{2}{c}{\textbf{Fine}}                                 \\ \cline{2-7} 
                                 & \multicolumn{1}{c}{\textbf{Relative Error↓}} & \textbf{Hit Rate↑} & \multicolumn{1}{c}{\textbf{Relative Error↓}} & \textbf{Accuracy↑} & \multicolumn{1}{c}{\textbf{Relative Error↓}} & \textbf{Accuracy↑} \\ \hline
Llama-3-8b                       & \hspace{0.1cm} 1.403±1.573**                                & 0.845              & \hspace{0.1cm} 0.834±1.298**                                & 0.385              & \hspace{0.1cm} 3.774±6.792**                                & 0.655              \\ \hline
GPT-3.5-turbo                    & \hspace{0.1cm} 1.071±1.181**                                & 0.850               & \hspace{0.1cm} 0.856±0.981**                                & 0.440              & \hspace{0.1cm} 1.568±3.856*                                 & 0.665              \\ \hline
Deepseek-v3                      &\hspace{0.1cm} 0.445±0.442*                                 & 0.850              & \hspace{0.1cm} 0.494±0.566                                  & 0.470              & \hspace{0.1cm} 1.055±1.562                                  & 0.660              \\ \hline
AgentCourt                       & \hspace{0.1cm} 0.464±0.513*                                 & 0.865              & \hspace{0.1cm} 0.493±0.496                                  & 0.500              & \hspace{0.1cm} 1.275±1.923*                                 & 0.670              \\ \hline
AgentsCourt                      & \hspace{0.1cm} 0.602±0.748**                                & 0.875              & \hspace{0.1cm} 0.601±0.617*                                 & 0.700              & \hspace{0.1cm} 1.973±5.286*                                 & 0.695              \\ \hline
PLJP                             & \hspace{0.1cm} 0.536±0.734**                                & 0.825              & \hspace{0.1cm} 0.523±0.605                                  & 0.690              & \hspace{0.1cm} 1.194±2.366                                  & 0.725              \\ \hline
SimCourt                         & \hspace{0.2cm}\textbf{0.350±0.367}                         & \textbf{0.880}     & \hspace{0.2cm}\textbf{0.410±0.435}                         & \textbf{0.730}     & \hspace{0.2cm}\textbf{0.770±1.210}                         & \textbf{0.835}     \\ \hline
\end{tabular}
\caption{Legal judgment prediction results comparing with baselines on imprisonment, probation and fine. Relative error (with standard deviation), hit rate and accuracy relative to the real documents are reported. The best-performing methods are highlighted with boldface. */** denotes that SimCourt performs significantly better than baselines at $p < 0.05/0.01$ level}
\label{CompareWithBaselines}

\end{table*}

\renewcommand{\arraystretch}{1.1}
\begin{table*}[]
\centering
\small
\begin{tabular}{l|cc|cc|cc}
\hline
\multicolumn{1}{c|}{\multirow{2}{*}{\textbf{Method}}} & \multicolumn{2}{c|}{\textbf{Imprisonment}}    & \multicolumn{2}{c|}{\textbf{Probation}}       & \multicolumn{2}{c}{\textbf{Fine}}             \\ \cline{2-7} 
\multicolumn{1}{c|}{}                                 & \textbf{Relative Error↓} & \textbf{Hit Rate↑} & \textbf{Relative Error↓} & \textbf{Accuracy↑} & \textbf{Relative Error↓} & \textbf{Accuracy↑} \\ \hline
SimCourt                                              & \textbf{0.350±0.367}     & \textbf{0.880}     & \textbf{0.410±0.435}     & \textbf{0.730}     & \textbf{0.770±1.210}     & \textbf{0.835}     \\ \hline
w/o Court Simulation                                  & 0.413±0.450              & 0.870              & 0.443±0.526              & 0.505              & 0.937±1.344              & 0.680              \\ \hline
w/o Memory                                            & 0.370±0.501              & 0.850              & 0.452±0.425              & 0.695              & 1.044±2.039              & 0.830              \\ \hline
w/o Strategy                                          & 0.424±0.497              & \textbf{0.880}     & 0.426±0.537              & 0.660              & 0.860±1.734              & 0.675              \\ \hline
\end{tabular}

\caption{Results of ablation study on legal judgment prediction task. Relative error (with standard deviation), hit rate and accuracy relative to the real documents are reported. The best-performing methods are highlighted with boldface. }
\label{AblationStudy}
\end{table*}






\textbf{Workflow}. Before the commencement of the court, each agent receive their basic courtroom tasks and fill their Profile module, which would be kept thorough the whole trial process. Subsequently, each agent initialize its own strategies. Prosecutor and Attonery initialize their Attack strategy and Defense strategy, while Defendant designs his Defense strategy and Judge formulates an Investigation strategy as guidance to investigate the details of the case. Specifically, we let each agent generate queries related to the application of law based on the case information, and then submit them to LegalOne to retrieve similar cases. We then let agents compile a list of relevant legal article titles and obtain their full content through the legal article retriever. The final Strategy module of an agent is built with the case information, its courtroom tasks, the retrieved legal provisions and similar cases, and the strategy types (i.e., Attack, Defense, and Investigation) assigned to its role.


As shown in Figure.\ref{fig:agent}, agents take the latest recordings as input and generate response with their Profile, Memory, and Strategy modules in prompt context. 
Then, at the end of each stage, we require each agent to conduct a reflection process before proceeding to the next stage, the workflow of which is also illustrated in Figure.\ref{fig:agent}. 
Specifically, they first adjust their Strategy module based on their short-term memory, long-term memory and retrieval results from tools. 
Then they update their long-term memory by summarizing their long-term memory together with the current short-term memory. 
After that, the short-term memory buffer is cleared and the agent is ready to participate the next stage in the trial. 









\begin{table}[h!]
{
\small
\begin{tabular}{cccc}
\Xhline{1.2pt}
\multicolumn{1}{c|}{Aspect}                               & Sim.             & Draw             & Hum.    \\ \hline\hline
\multicolumn{4}{c}{\textbf{Trial Investigation-Judge}}                                         \\ \hline
\multicolumn{1}{c|}{Clarity of Questioning Structure}   & \textbf{0.70} & 0.30          & 0.00 \\
\multicolumn{1}{c|}{Neutrality and Procedural Control}  & \textbf{0.85} & 0.10          & 0.05 \\
\multicolumn{1}{c|}{Professional Evidence Examination}     & \textbf{0.80} & 0.15          & 0.05 \\ \hline
\multicolumn{4}{c}{\textbf{Trial Investigation-Prosecutor}}                                    \\ \hline
\multicolumn{1}{c|}{Proper Questioning Strategy}    & \textbf{0.90} & 0.05          & 0.05 \\
\multicolumn{1}{c|}{Precise Legal Terminology}    & \textbf{1.00} & 0.00          & 0.00 \\
\multicolumn{1}{c|}{Lawful Prosecutorial Questioning}   & 0.40          & \textbf{0.45} & 0.15 \\ \hline
\multicolumn{4}{c}{\textbf{Trial Investigation-Attorney}}                                      \\ \hline
\multicolumn{1}{c|}{Relevance and Precision in Questioning}         & \textbf{1.00} & 0.00          & 0.00 \\
\multicolumn{1}{c|}{Awareness of Procedural Legitimacy}     & \textbf{0.95} & 0.05          & 0.00 \\
\multicolumn{1}{c|}{Awareness of Defendant’s Rights}      & \textbf{1.00} & 0.00          & 0.00 \\ \hline \hline
\multicolumn{4}{c}{\textbf{Evidence Presentation-Judge}}                                       \\ \hline 
\multicolumn{1}{c|}{Normative Conduct}                  & \textbf{0.85} & 0.15          & 0.00 \\
\multicolumn{1}{c|}{Cross-Exam Legality Control}     & \textbf{0.55} & 0.30          & 0.15 \\
\multicolumn{1}{c|}{Awareness of Fair Trial Safeguards} & \textbf{0.65} & 0.30          & 0.05 \\ \hline
\multicolumn{4}{c}{\textbf{Evidence Presentation-Prosecutor}}                                  \\ \hline
\multicolumn{1}{c|}{Accuracy in Evidence Presentation}  & \textbf{0.90} & 0.10          & 0.00 \\
\multicolumn{1}{c|}{Moderation in Aggressive Advocacy}  & \textbf{0.90} & 0.10          & 0.00 \\
\multicolumn{1}{c|}{Proper Response to Objections}      & \textbf{0.95} & 0.05          & 0.00 \\ \hline
\multicolumn{4}{c}{\textbf{Evidence Presentation-Attorney}}                                    \\ \hline
\multicolumn{1}{c|}{Precision in Challenging Key Issues}    & \textbf{0.95} & 0.05          & 0.00 \\
\multicolumn{1}{c|}{Rigor in Evidence Analysis}         & \textbf{0.95} & 0.05          & 0.00 \\
\multicolumn{1}{c|}{Effectiveness in Evidence Rebuttal}        & \textbf{1.00} & 0.00          & 0.00 \\ \hline\hline
\multicolumn{4}{c}{\textbf{Trial Debate-Judge}}                                                \\ \hline
\multicolumn{1}{c|}{Clear Adversarial Framing}    & \textbf{0.85} & 0.10          & 0.05 \\
\multicolumn{1}{c|}{Impartial Verbal Interventions}     & \textbf{0.65} & 0.25          & 0.10 \\
\multicolumn{1}{c|}{Pace and Order Control}             & \textbf{0.85} & 0.10          & 0.05 \\ \hline
\multicolumn{4}{c}{\textbf{Trial Debate-Prosecutor}}                                           \\ \hline
\multicolumn{1}{c|}{Logical Coherence of Accusation}    & \textbf{0.95} & 0.00          & 0.05 \\
\multicolumn{1}{c|}{Accuracy in Legal Citation}         & \textbf{1.00} & 0.00          & 0.00 \\
\multicolumn{1}{c|}{Effective Defense Rebuttal}        & \textbf{0.95} & 0.05          & 0.00 \\ \hline
\multicolumn{4}{c}{\textbf{Trial Debate-Attorney}}                                             \\ \hline
\multicolumn{1}{c|}{Clarity of Defense Arguments}       & \textbf{0.95} & 0.00          & 0.05 \\
\multicolumn{1}{c|}{Logical Rigor in Legal Reasoning}   & \textbf{0.90} & 0.05          & 0.05 \\
\multicolumn{1}{c|}{Balanced Legal and Emotional Appeal}         & \textbf{0.85} & 0.10          & 0.05 \\ \hline\hline
\multicolumn{4}{c}{\textbf{Overall Performance}}                                               \\ \hline
\multicolumn{1}{c|}{Judge}                              & \textbf{0.95} & 0.00          & 0.05 \\ \hline
\multicolumn{1}{c|}{Prosecutor}                         & \textbf{1.00} & 0.00          & 0.00 \\ \hline
\multicolumn{1}{c|}{Attorney}                           & \textbf{1.00} & 0.00          & 0.00 \\ \Xhline{1.2pt}

\end{tabular}
}
\caption{The results of trial process evaluation, which represents annotators' preferences on 30 aspects across 20 cases. ``Sim." represents proportions of cases where annotators prefer SimCourt, ``Hum." represents proportions where annotators prefer the real ones created by human, and `Draw' indicates that annotators cannot distinguish which is better.}
\label{process_table}
\end{table} 

\section{Evaluation Framework and Benchmark}

For a comprehensive evaluation for LLM-based mock trial, we develop an evaluation framework as well as a benchmark based on both the objective goals of the court and a widely used process evaluation guideline. The evaluation framework consists of two parts: judgment prediction evaluation and simulation process evaluation.

\paragraph*{Judgment Prediction Evaluation}

For judgment prediction, we extract the final judgment made by Judge in the simulation and compare it with the ground truth recorded in the original case document on 3 aspects, i.e., imprisonment, probation and fine, from two perspectives, i.e., categorical accuracy and quantitative error. 

\textbf{Categorical Accuracy} focuses on evaluating whether the judgment produced by SimCourt fall into the correct categories supported by law.
For \textit{imprisonment} prediction, inspired by the LJP track in CAIL2018 \cite{xiao2018cail2018largescalelegaldataset}, we examine whether the prediction imprisonment number is within the same acceptable range with the ground truth value from the case document, which we define a ``hit''. Specifically, if and only if the predicted imprisonment and the actual imprisonment both fall within the same statutory sentencing range, the prediction is considered to be a ``hit''. Accordingly, hit rate is defined as the proportion of cases where the predicted sentence hits the correct statutory sentencing interval, which can be calculated as the followings: 
\begin{equation}
\small
Hit \space Rate=\frac{1}{N} \sum_{i=1}^{N} \mathds{1}[L_i\leq P_i \leq U_i],
\end{equation}
where $N$ denotes the number of cases. For the $i$th case, $P_i$ represents the predicted imprisonment while $L_i$ and $U_i$ denotes the lower and upper bounds of the sentencing interval of actual imprisonment, respectively. For \textit{probation} prediction, we treat it as a binary classification task. A prediction is considered correct if and only if both the predicted and the actual judgment agree that probation is applicable, or both agree that it is not applicable. \textit{Fine} prediction is evaluated in the same manner with probation.

\textbf{Quantitative Error} evaluates the relative error between the predicted judgment by court simulation and the ground truth from the case document, which provides a different angle to understand the performance of judgment prediction.
For example, let the ground truth imprisonment for the $i$th case be $T_i$. 
Then the relative error of predicted imprisonment is computed as
\begin{equation}
 \small
 Relative \space Error=\frac{1}{N} \sum_{i=1}^{N} \frac{\left | P_i-T_i \right | }{T_i},
\end{equation}
The lower the relative error is, the better the imprisonment prediction is.
The relative error for probation and fine is computed in a similar way.
Note that in the above equation, we explicitly exclude cases where $T_i$ is 0 or where $T_i$ represents a non-numeric sentence (e.g., the death penalty) to avoid numerical problems. This exclusion may introduce a certain bias to the metric.
Therefore, a better practice is to jointly consider both the categorical accuracy and quantitative error for the evaluation of  different models or systems.


\paragraph*{Trial Process Evaluation} Based on the evaluation guideline for Chinese legal professionals, we propose a process evaluation framework which lies on 30 different aspects, as shown in Table.\ref{process_table}. Here we focus the evaluation on three major roles and three most important stages in the court, namely Judge, Prosecutor, Attorney, and Trial Investigation, Evidence Presentation, Trial Debate, respectively. In addition to the 27 stage-wise evaluation, we also exams the overall performance of Judge, Prosecutor, and Attorney. In practice, quantifying the actual performance of each role in each court stage is usually difficult even for legal experts.
Therefore, we conduct pairwise evaluation instead of pointwise evaluation for better reproducibility.
Specifically, we collect real recording of trials in Chinese court and use the same case materials to conduct court simulation with SimCourt.
Then we hire human experts to compare the simulated trial with the real trial and annotate whether the roles in the simulation outperform, underperform, or perform equally with those in the real trial on each of the 30 aspects.


\paragraph*{Benchmark Construction}

To ensure the reproducibility of our work and support future studies, we construct and release a mock trial benchmark based on our proposed evaluation framework. For judgment prediction evaluation, we need the case description as input and the actual judgment as ground truth, which can be extracted from case documents automatically.
Thus, we randomly selected 200 cases from LeCaRDv2 \cite{li2024lecardv2}, a large-scale Chinese legal case retrieval dataset containing 55,192 cases, from 40 most common charges (5 per charge). 
To avoid unnecessary confusion in case descriptions, all selected cases are first-instance trials involving a single defendant. We used Deepseek-v3 model \cite{liu2024deepseek} to extract the defendant’s information, indictment, and evidence as input, and refined details with human legal expert review. 

For process evaluation, the input should be the case descriptions and the recordings of real trials in court.
To this end, we downloaded 20 open trial videos with the full trial records provided on China Court Trial Online and converted the audio into text (with human proofreading). The sampled recordings cover 20 distinct charges drawn from the same 40-charge set used in judgment prediction evaluation. The recordings are 37-minutes long and contain 8,000 words on average. We manually extracted the indictment, defendant’s information, and corresponding evidence to serve as input for court simulation. 

Note that none of the selected cases above is included in the corpus of LegalOne, which means that no model can directly get the answers from retrieved case documents.

\section{Experiments}



\paragraph*{Legal Judgment Prediction}
This section presents experiments on legal judgment prediction (LJP) tasks, focusing on the effectiveness of imprisonment, probation, and fine decisions in the predicted documents relative to the real ones.
%





\textbf{Comparison with Baselines}.
First, we examine the performance of different vanilla models, i.e. Llama-3-8b \cite{dubey2024llama}, GPT-3.5-turbo \cite{brown2020language}, and Deepseek-v3 \cite{liu2024deepseek}, the task of which is directly predicting the judgment based on the case information. Afterwards, we introduce another 3 state-of-the-art baseline methods: the aforementioned AgentCourt \cite{chen2024agentcourt} and AgentsCourt \cite{he2024agentscourt} as well as PLJP \cite{wu2023precedent}, which is a LLM-based LJP method via case retrieval and detailed analysis. For AgentCourt and AgentsCourt, the debating round is set to 3, which is the same as SimCourt. Besides, for fair comparison, we choose Deepseek-v3 model as the base model for SimCourt and the 3 baseline methods. We also provide all those baselines with the same tools in SimCourt, i.e. the legal article retriever and LegalOne. 



The results are listed in Table.\ref{CompareWithBaselines}. SimCourt outperforms all baselines on every aspect and metric, with the lowest error rate and standard deviation as well as the highest hit rate and accuracy. This results from the adequacy and rationality of process simulation. Besides, we observe that other simplified court simulation methods, i.e., AgentCourt, AgentsCourt, and the non-simulation LJP method, i.e., PLJP, fail to consistently outperform vanilla methods across the six metrics. This phenomenon indicates that insufficient court interactions and unprofessional legal agents can instead confuse the judge, leading to greater numerical deviations, further highlighting the effectiveness of our process simulation and agent design. It is also notable that Deepseek-v3 model performs best among the vanilla models, demonstrating the fundamental capability of the chosen base model.

\textbf{Ablation Study}. To further evaluate the effectiveness of our framework, we implement ablation study on trial process and agent modules. Results are listed in Table.\ref{AblationStudy}. We can find that SimCourt performs best among all aspects and metrics, proving the effectiveness of our framework design. Without court simulation, the hit rate and accuracy declines, while the relative error increase, highlighting the necessity of simulating judicial proceedings. Moreover, removing either the Memory module or the Strategy module leads to performance degradation across most metrics, highlighting the necessity of both modules in the system design. 



\begin{figure*}
    \centering
    \includegraphics[width=1\linewidth]{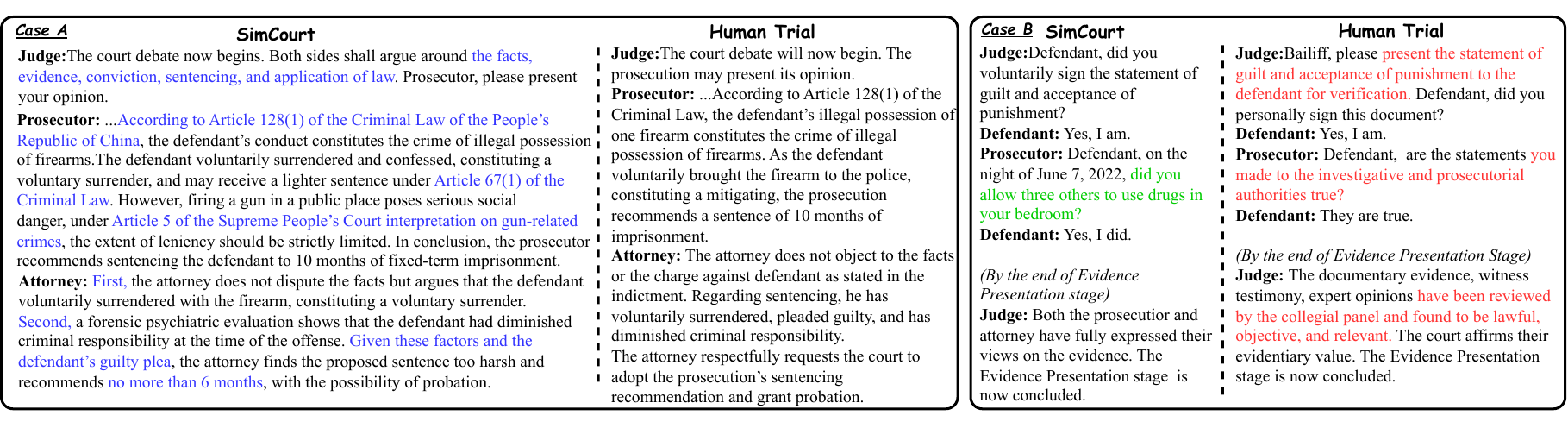}
    \caption{Records comparison between SimCourt and human trial.
Case A showcases typical strengths of SimCourt, with effective expressions highlighted in blue.
Case B illustrates typical weaknesses, where inappropriate statements are highlighted in green, and well-performed statements from the human trial are highlighted in red.}
    \label{case_study}
\end{figure*}


\paragraph*{Trial Process Evaluation}

Based on the process evaluation framework and benchmark, we applied trial process evaluation using human preference annotations. Case study is also applied to directly presents the strengths and weaknesses of our simulation.


\textbf{Human Annotation}. We invited 3 legal experts to annotate the quality of the simulation process. Each annotator has passed the National Unified Legal Professional Qualification Examination. We provided annotators with the 20 pairs of simulated and real trial records, with the order of the two records within each pair randomly shuffled to ensure a blind evaluation setting. For each pair, annotators were asked to indicate their preference across 30 evaluation aspects by selecting which record (the first or the second) is better, or by choosing ``draw'' if the two were indistinguishable. When annotators judged the simulated record as better or selected ``draw'', it indicated that the simulated record was not inferior to the real one and had reached a satisfactory level. Under this circumstance, the average Kappa value among annotators is 0.833, reflecting a high level of agreement and demonstrating the validity and reliability of the annotation. 

Afterwards, for each case, we aggregated the 3 annotations on each aspect to derive a final preference. Specifically, we excluded “draw” judgments and determined the majority preference (i.e., simulation or human) for each aspect. If no majority exists, the result is recorded as a draw. Finally, we summarized the evaluation result in Table.\ref{process_table}.

\textbf{Results and Analysis}. Overall, SimCourt demonstrates strong performance across the majority of evaluation aspects. For Judge, the simulated agents are preferred in most cases, particularly in procedural control (85\%), adversarial framing (85\%), and clarity of questioning structure (70\%), reaching an overall preference rate of 95\%. For Prosecutor, the system achieves consistently high scores, with 100\% preference in precise legal terminology and accurate legal citation, and 95\% in both logical coherence of accusations and effectiveness of rebuttals. The defense attorney also shows outstanding performance, with over 95\% preference rates in evidence rebuttal, legal reasoning, and rights protection, and 100\% preference in rebuttal effectiveness and awareness of defendant’s rights. These strengths benefit from the well-structured simulation process and the carefully designed agent framework. The results further suggest that with appropriate integration of general-domain and legal-domain knowledge, LLM-based agents can even outperform human legal practitioners in certain tasks.

However, several limitations remain. For Judge, relatively lower preference lies in cross-examination legality control (55\%) and impartial verbal interventions (65\%). This may result from limited flexibility in handling adversarial exchanges and limited control over rhetorical tone in dynamic courtroom scenarios. For Prosecutor, this role underperforms in lawful questioning, with only 40\% of cases rated better than human performance. This indicates that the agent may lack nuanced understanding of procedural legitimacy, such as restrictions on leading questions, adherence to evidence sequence, and safeguarding of defendant rights. Attorney generally performs well, only with the preference rate for balanced legal and emotional appeal slightly lower (85\%). This suggests that current LLMs still face challenges in generating affective, emotionally resonant arguments.

\textbf{Case Study}. To directly present the strengths and weaknesses of the simulation process, we selected two typical cases, as shown in Figure.\ref{case_study}. Case A shows typical strengths of SimCourt while Case B demonstrates typical weaknesses.


In \textit{Case A}, the simulated judge emphasized that the debate should focus on key issues such as sentencing, offering a clearer adversarial framing than the human judge. The simulated prosecutor also demonstrated stronger legal grounding by citing multiple legal articles with detailed explanations, whereas the human prosecutor referenced only one article with minimal elaboration. Additionally, the SimCourt attorney presented a logical and well-reasoned defense, outperforming the human attorney who made a relative simpler defense statement with no sentence recommendation.

In \textit{Case B}, the human judge instructed the bailiff to return the statement to the defendant for verification and formally acknowledged the evidentiary validity, demonstrating proper legal procedure. In contrast, the simulated judge failed to recognize or apply basic principles of legality control. Furthermore, the human prosecutor questioned the defendant regarding the authenticity of the statements, whereas the simulated prosecutor posed leading questions, resulting in legally flawed prosecutorial conduct. Hopefully, through training from real courtroom records, simulated agents may further improve their legal knowledge and procedural awareness, helping to mitigate current limitations.

\section{Conclusion}

In this paper, we introduce SimCourt, a court simulation framework based on real Chinese trial procedures. A specially designed agent framework are proposed, helping agents simulate professional practitioners. Evaluation framework and benchmark on both legal judgment prediction and simulation process evaluation are developed. Further experiments and analysis demonstrate the potential application value of our framework and highlight directions for future improvement.

\bibliography{reference}

\end{document}